\documentclass[aps,superscriptaddress,prc,twocolumn,nofootinbib]{revtex4}

\usepackage{graphicx}
\usepackage{epstopdf}
\usepackage{subfigure}
\usepackage{epsfig}
\usepackage{amsmath,amssymb,amsfonts}
\usepackage{color}
\usepackage[utf8]{inputenc}

\begin{document}

\title{Gluon contribution to open heavy-meson production in heavy-ion collisions }

\author{Shanshan Cao}
\affiliation{Nuclear Science Division, Lawrence Berkeley National Laboratory, Berkeley, CA 94720, USA}
\author{Guang-You Qin}
\affiliation{Institute of Particle Physics and Key Laboratory of Quark and Lepton Physics (MOE), Central China Normal University, Wuhan, 430079, China}
\author{Xin-Nian Wang}
\affiliation{Nuclear Science Division, Lawrence Berkeley National Laboratory, Berkeley, CA 94720, USA}
\affiliation{Institute of Particle Physics and Key Laboratory of Quark and Lepton Physics (MOE), Central China Normal University, Wuhan, 430079, China}

\date{\today}


\begin{abstract}
A sizable contribution to heavy quark production in high-energy hadronic and nuclear collisions comes from heavy quark-antiquark pair production from gluon splitting during the parton shower evolution.  We investigate the effect of gluon-medium interaction on open heavy flavor spectra in ultra-relativistic heavy-ion collisions. The interaction of hard gluons and heavy quarks with the hot QCD medium is simulated by utilizing a Langevin transport model that simultaneously incorporates contributions from collisional and radiative processes. It is found that while the gluon splitting channel has quite an important contribution to the single $D$ meson production cross section, its influence on the final heavy meson nuclear modification turns out to be quite modest because the average lifetime of hard gluons is short before splitting into heavy quark-antiquark pairs during the evolution and propagation of the parton shower.
\end{abstract}

\maketitle



{\em Introduction --}
Heavy quarks are regarded as one of the valuable tools to probe the strongly interacting quark-gluon plasma (sQGP) produced in ultra-relativistic heavy-ion collisions. Among the many interesting measurements of heavy flavors at the Relativistic Heavy-Ion Collider (RHIC) and the Large Hadron Collider (LHC) are observations of the strong nuclear modification and the large azimuthal anisotropy for heavy flavor hadrons and their decay electrons, which are comparable to those of light hadrons \cite{Adare:2010de,Adare:2014rly,Tlusty:2012ix,Adamczyk:2014uip,Grelli:2012yv,Abelev:2013lca}. These results provide a big challenge to our naive expectation about the mass hierarchy of parton energy loss, i.e., heavy quarks tend to lose less energy than light flavor partons in dense nuclear matter due to the supression of the phase space for collinear gluon radiation by their large masses. There have been various transport models established to investigate the evolution dynamics of heavy quarks inside the hot and dense QGP, such as parton cascade models \cite{Molnar:2006ci,Zhang:2005ni,Uphoff:2011ad,Uphoff:2012gb} based on the Boltzmann transport approach , the linear Boltzmann approach coupled to a hydrodynamic background \cite{Gossiaux:2010yx,Nahrgang:2013saa}, the Langevin-based approaches \cite{Moore:2004tg,He:2011qa,Young:2011ug,Cao:2011et,Cao:2012jt,Cao:2013ita,Cao:2015hia,Cao:2015cba} and the parton-hadron-string-dynamics transport approach \cite{Song:2015sfa}. For a recent review of heavy flavor production in relativistic nuclear collisions,  readers are referred to Ref. \cite{Averbeck:2015jja}.

In most current descriptions of heavy quark dynamics in a dense nuclear medium, one common practice is to assume that the majority of heavy quarks are produced in the primordial stage of heavy-ion collisions and they probe the entire evolution history of the QGP fireball. This may be well justified at the leading order where heavy quarks are produced in hard scattering processes via the pair production by the light parton fusion ($gg\rightarrow c\bar{c}$, $q\bar{q}\rightarrow c\bar{c}$) and the flavor excitation ($gc\rightarrow gc$, $qc\rightarrow qc$). As one goes beyond the leading-order approximation, the contribution to heavy quark production from the splitting of hard virtual gluons ($g\rightarrow c\bar{c}$) has to be considered. 
It has been pointed out in Refs. \cite{Bedjidian:2004gd,Norrbin:2000zc,Aad:2012ma,Huang:2013vaa,Huang:2015mva} that the contribution from the gluon splitting process to heavy quark production is equally important as compared to contributions from the pair production and flavor excitation channels. Therefore, in ultra-relativistic heavy-ion collisions, one needs to consider the medium effect on these hard virtual gluons before they split into heavy quark-antiquark pairs. We note that in Ref. \cite{Huang:2013vaa,Huang:2015mva}, it is argued that the inclusion of gluon splitting is very important to describe the observed suppression for $b$-tagged jets in heavy-ion collisions. 

In this work, we re-investigate the contribution from the gluon splitting process to heavy flavor production in relativistic heavy-ion collisions. In particular, the influence of a gluon's energy loss before its splitting into a heavy quark-antiquark pair on the nuclear modification of the final open heavy meson spectra is studied in details. We note that one of the key factors here is the lifetime of hard gluons before they split into heavy quark-antiquark pairs. If the gluon's lifetime is shorter than the formation time of the QGP, then the gluon splitting process is just a modification of the initial heavy quark spectra in addition to the conventional pair production and the flavor excitation channels. However, if the gluon lifetime is sufficiently large, one has to include the interaction of hard gluons with the hot and dense QGP before they split into heavy quark-antiquark pairs. In this case, since gluons tend to lose more energy than heavy quarks, the inclusion of the gluon splitting channel and gluon energy loss before the splitting might result in stronger nuclear modification for the final heavy flavor mesons. 

For our purpose, here we use \textsc{Pythia} 6 \cite{Sjostrand:2006za} to simulate the gluon splitting process in vacuum and extract the splitting probability of hard gluons, the gluon lifetime, and the gluon to heavy quark-antiquark pair splitting function. The medium effect on the gluon splitting process is simulated via our Langevin transport model as developed in the earlier works where the effect of both collisional and radiative processes are incorporated simultaneously \cite{Cao:2013ita,Cao:2015hia}. We show that due to the short lifetime of hard gluons, the energy loss experienced by hard gluons before splitting into heavy quark-antiquark pairs is not large enough to significantly affect the nuclear modification of the transverse momentum spectra of single heavy flavor mesons.

\begin{figure}[tb]
  \epsfig{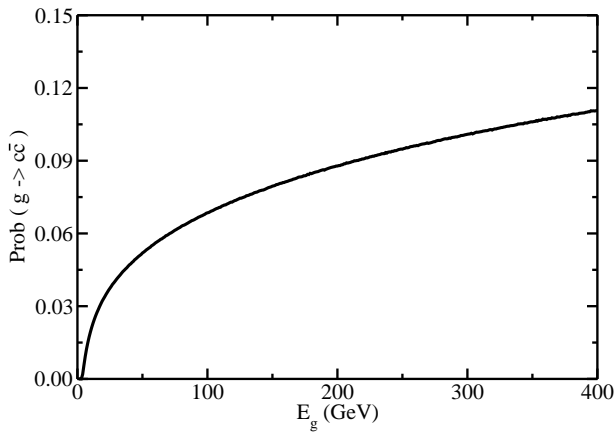}
  \caption{(Color online) Probability for gluons to split into $c\bar{c}$ as a function of gluon energy.}
 \label{fig:plot-prob_split}
\end{figure}

{\em Results --}
We first show in Fig.~\ref{fig:plot-prob_split} the probability for hard gluons to split into $c\bar{c}$ pairs as a function of gluon energy. This probability is obtained from a \textsc{Pythia} simulation by setting the maximum allowed mass for each parton ($Q_{\rm max}$) in the parton shower (\textsc{pyshow}) as the initial gluon energy. As seen in Fig.~\ref{fig:plot-prob_split}, the gluon splitting probability increases with gluon energy. For gluons with $400$~GeV energy, the probability to split into heavy quark-antiquark pairs can exceed 10\%.

\begin{figure}[tb]
  \epsfig{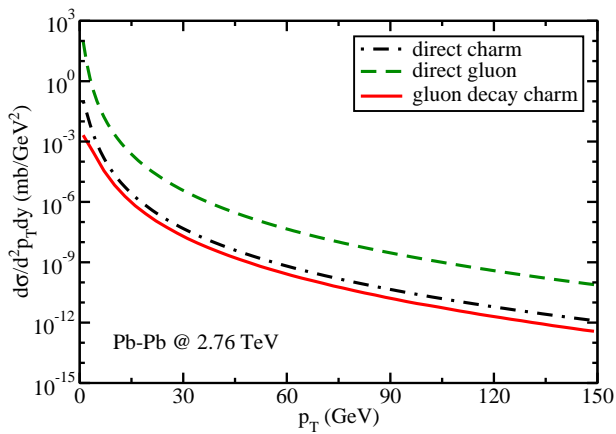}
  \caption{(Color online) Comparison of initial transverse momentum spectra for charm quarks from direct (pair production and flavor excitation) channels, the initial hard gluons, and charm quarks from gluon splitting channel for Pb-Pb collisions at 2.76~TeV.}
 \label{fig:plot-initialSpectra}
\end{figure}

Using the above splitting probability, together with the splitting function $P_{g \to c\bar{c}}(x,k_\perp^2)$, where $x$ is the factional of energy and $k_\perp$ is the transverse momentum of the heavy quark with respect to the initial hard gluon, one may obtain the spectrum of the produced charm quarks from the initial gluon spectra. 
We use the same \textsc{Pythia} simulation to extract the gluon splitting function $P_{g\to c\bar{c}}(x, k_\perp^2)$. In Fig.~\ref{fig:plot-initialSpectra}, we show the spectrum of charm quarks from the gluon splitting process ($g\rightarrow c\bar{c}$) and compare it with that from the direct production channels, i.e., pair production and flavor excitation. In our work, the direct charm and direct gluon spectra are obtained with leading-order perturbative QCD calculation in which the CTEQ parametrization of parton distribution functions \cite{Lai:1999wy} is adopted and the EPS09 parametrizations \cite{Eskola:2009uj} are utilized for the nuclear shadowing effect in nucleus-nucleus collisions. We see an important contribution of the gluon-splitting channel to charm quark production in Fig.~\ref{fig:plot-initialSpectra}, similar to the finding in Refs.~\cite{Bedjidian:2004gd,Norrbin:2000zc,Aad:2012ma,Huang:2013vaa}.

\begin{figure}[tb]
  \epsfig{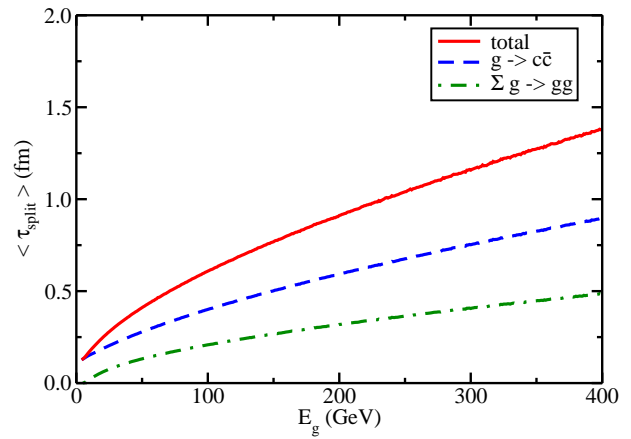}
  \caption{(Color online) Average splitting time for $g \to c\bar{c}$ as a function of the initial gluon energy.}
 \label{fig:plot-tau-avr}
\end{figure}

\begin{figure}[tb]
  \epsfig{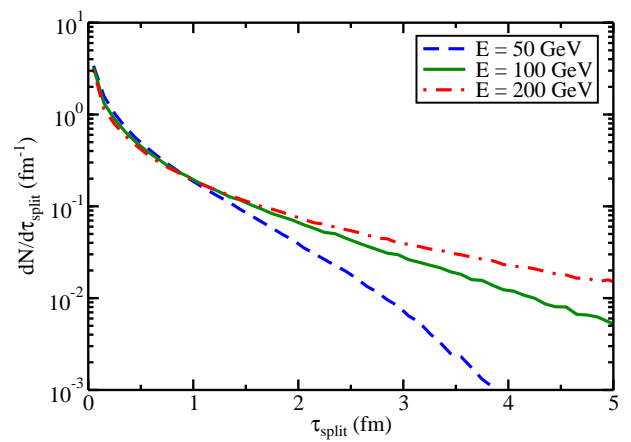}
  \caption{(Color online) Probability distribution of the $g \to c\bar{c}$ splitting time for three different gluon energies.}
 \label{fig:plot-tau-distr}
\end{figure}

To investigate the medium effect on heavy quarks produced from the gluon splitting channel, one has to consider the interaction of gluons with the dense nuclear medium before splitting into heavy quark-antiquark pairs. One of the key quantities that determines the amount of medium effect on hard gluons is the gluon lifetime $\tau_\mathrm{split}=2E/Q^2$, which is related to its virtuality $Q^2$ and energy $E$. We utilize a similar method employed by Ref. \cite{Eskola:1993cz} and keep track of the full parton shower chain initialized by a hard gluon to obtain the gluon lifetime in our \textsc{Pythia} simulations. Here we identify the gluon lifetime as the accumulated time before the splitting to a heavy quark-antiquark pair from an initial hard gluon, which is obtained by summing up the splitting times for all subprocesses in the shower chain. In Fig.~\ref{fig:plot-tau-avr}, we show the average splitting time of a hard gluon as a function of the gluon's initial energy. Contributions from the final splitting process $g\rightarrow c\bar{c}$ and that from the previous $g\rightarrow gg$ chain are shown separately. One can see that it takes longer time for more energetic gluons to split into final $c\bar{c}$ pairs. In Fig.~\ref{fig:plot-tau-distr}, we show the probability distribution of the gluon lifetime (splitting time) for three different initial gluon energies. We can see that most hard gluons split at very early times of heavy-ion collisions, even before the formation (thermalization) of the QGP (at $\tau_0=0.6$~fm). However, there still exists some fraction of gluons that survive the pre-equilibrium phase and interact with the dense QGP matter before splitting into $c\bar{c}$ pairs. We also see that the number of long-lived gluons increases with the gluon's initial energy.

To simulate the dynamical evolution of hard partons inside the QGP, we adopt the following modified Langevin equation 
developed in Refs.~ \cite{Cao:2013ita,Cao:2015hia}:
\begin{equation}
\label{eq:modifiedLangevin}
\frac{d\vec{p}}{dt}=-\eta_D(p)\vec{p}+\vec{\xi}+\vec{f}_g .
\end{equation}
In the above equation, the first two terms on the righthand side denote the drag force and the thermal random force exerted on the hard parton while it diffuses inside a thermal medium, and the third term is introduced to describe the recoil force that the parton experiences via radiating gluons. 
Here the interaction of heavy quarks and gluons with the medium is simulated in the same Langevin framework. For a minimal setup, we assume that $\vec{\xi}$ is independent of the parton's momentum and virtuality and satisfies the correlation function of a white noise $\langle\xi^i(t)\xi^j(t')\rangle=\kappa\delta^{ij}\delta(t-t')$, where $\kappa$ represents the coefficient of momentum diffusion  and is related to the spatial diffusion coefficient $D$ via $D=2T^2/\kappa$. The diffusion coefficients of quarks and gluons differ by their color factors: $\kappa_Q/\kappa_G=C_F/C_A$. For the medium-induced gluon bremsstrahlung, the radiation probability within the time interval $[t,t+\Delta t]$ is determined by utilizing the average number of radiated gluons,
\begin{equation}
\label{eq:gluonnumber}
P_\mathrm{rad}(t,\Delta t)=\langle N_\mathrm{g}(t,\Delta t)\rangle = \Delta t \int dxdk_\perp^2 \frac{dN_\mathrm{g}}{dx dk_\perp^2 dt}.
\end{equation}
In the simulation, we choose sufficiently small time step $\Delta t$ to guarantee $P_{\rm rad}(t, \Delta t) < 1$. 
The gluon distribution function in our calculation is adopted from the higher-twist calculation \cite{Guo:2000nz,Majumder:2009ge,Zhang:2003wk},
\begin{align}
\frac{dN_\mathrm{g}}{dx dk_\perp^2 dt}=\frac{2\alpha_s C_A  P(x)}{\pi k_\perp^4} \hat{q}
 \left(\frac{k_\perp^2}{k_\perp^2+x^2 M^2}\right)^4 {\sin}^2\left(\frac{t-t_i}{2\tau_f}\right),\notag\\
 & \label{eq:gluondistribution}
\end{align}
where $x$ is the fractional energy taken away by the emitted gluon from its parent hard parton, $k_\perp$ is its transverse momentum, $P(x)$ is the splitting function for a given process, and $\tau_f={2Ex(1-x)}/{(k_\perp^2+x^2M^2)}$ is the formation time of the radiated gluon, where $E$ and $M$ are the energy and mass of the parent parton, respectively. We set $M=1.27$~GeV for charm quarks when calculating their in-medium transport. In the above equation, $\hat{q}$ is the parent parton's transport coefficient and is related to its momentum space diffusion coefficient $\kappa$ as: $\hat{q}=2\kappa$. We note that in our transport model there is only one free parameter which we choose to be the spatial diffusion coefficient $D$ of heavy quarks. In the following calculation, it is fixed to be $D=5/(2\pi T)$ which provides the best description of $D$ meson observables at the LHC and RHIC where only the direct channels for charm quark production are considered \cite{Cao:2015hia}. 

\begin{figure}[tb]
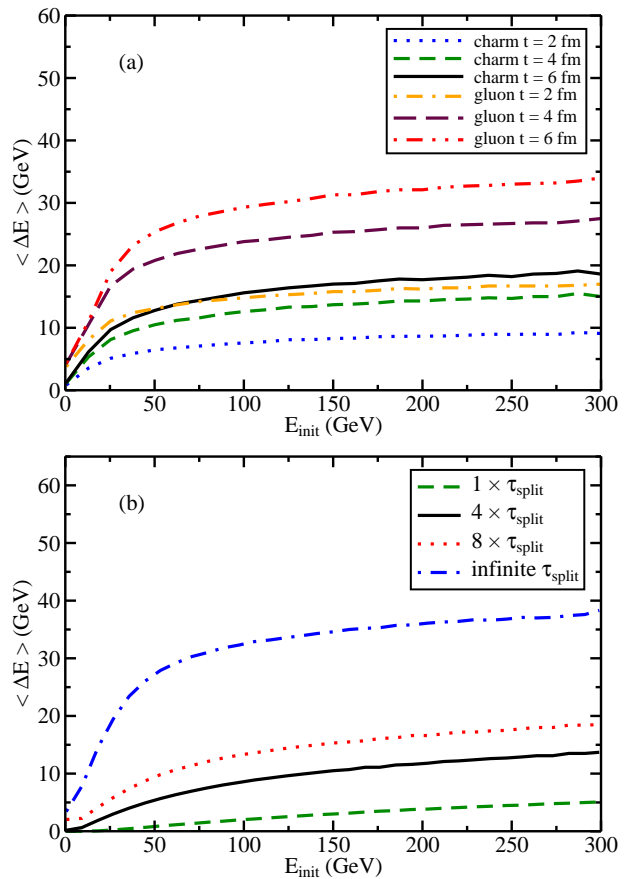

 \subfigure{\label{fig:plot-eLoss-tot}
  \epsfig{file=plot-eLoss-tot.eps, width=0.45\textwidth, clip=}}
 \subfigure{\label{fig:plot-GeLoss}
  \epsfig{file=plot-GeLoss.eps, width=0.45\textwidth, clip=}}
  \caption{(Color online) (a) Energy loss of gluons and charm quarks in a QGP medium for three evolution times. (b) Energy loss experienced by gluons before splitting into $c\bar{c}$ for different splitting times.}
\end{figure}

In Fig. \ref{fig:plot-eLoss-tot}, we compare the energy loss of hard gluons and charm quarks as a function of their initial energy at different evolution times in the hot QGP medium created in central Pb-Pb collisions at the LHC. Here we utilize a (2+1)-dimensional viscous hydrodynamic model \cite{Qiu:2011hf} to obtain the spacetime evolution profiles of the QGP fireball. As shown in the figure, the energy loss experienced by hard gluons is approximately twice of that by charm quarks given the same propagation time inside the medium. This is mostly due to different color factors of heavy quarks and gluons. In Fig.~\ref{fig:plot-GeLoss}, we show the energy loss experienced by a hard gluon for different propagation time before splitting into heavy quark-antiquark pairs. We can see that with a realistic splitting time as obtained from \textsc{Pythia}, a 300~GeV hard gluon only loses about 5~GeV energy before splitting into charm quark pairs, which is much smaller than the case that the splitting process happens after the gluon travels outside the QGP matter (as shown by the infinite splitting time curve). 

\begin{figure}[tb]
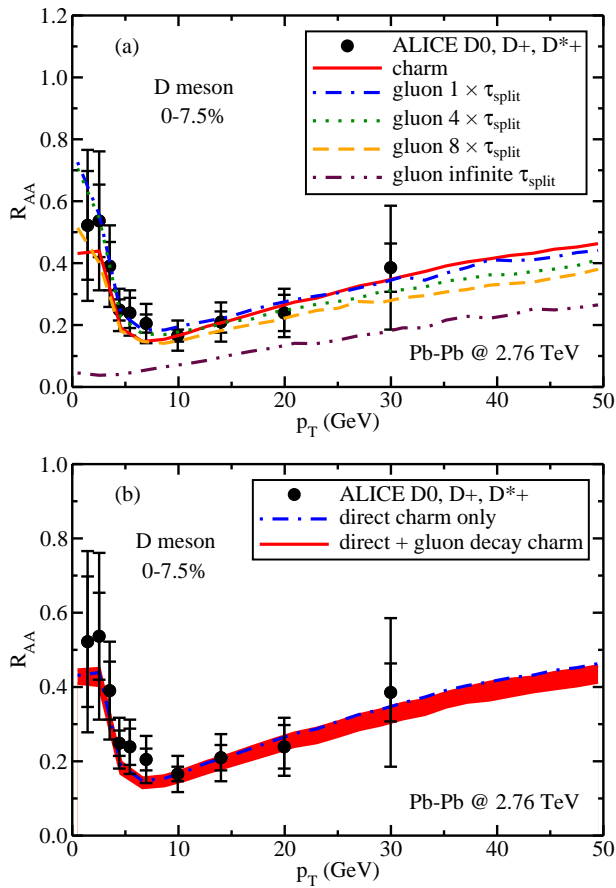

 \subfigure{\label{fig:plot-RAA_DD5-LHC-0-7d5-1}
      \epsfig{file=plot-RAA_DD5-LHC-0-7d5-1.eps, width=0.45\textwidth, clip=}}
 \subfigure{\label{fig:plot-RAA_DD5-LHC-0-7d5-2}
      \epsfig{file=plot-RAA_DD5-LHC-0-7d5-2.eps, width=0.45\textwidth, clip=}}
 \caption{(Color online) Comparison of $D$ meson nuclear modification factors $R_\mathrm{AA}$ in central Pb-Pb collisions: (a) direct and gluon-splitting channels with different gluon lifetimes, (b) with and without the inclusion of the gluon-splitting channel.}
 \label{fig:plot-RAA_DD5-LHC-0-7d5}
\end{figure}

Finally, we present our results for the final $D$ meson nuclear modification factor $R_\mathrm{AA}$ for 0-7.5\% central Pb-Pb collisions at the LHC. The hadronization of heavy quarks into the color-neutral bound states is simulated following Ref.~\cite{Cao:2015hia} via a hybrid model of fragmentation plus coalescence mechanisms. The effect of $D$ meson scatterings inside the hadron gas on its $R_\mathrm{AA}$ has been investigated in our previous study \cite{Cao:2015hia} and shown to be small, and thus is neglected in this work. In Fig.~\ref{fig:plot-RAA_DD5-LHC-0-7d5-1}, we first compare the nuclear modification factor for $D$ mesons produced from the direct channels and those from the gluon splitting channel. We can see that using the realistic splitting time $\tau_\mathrm{split}$ obtained from \textsc{Pythia} produces very small difference for the $D$ meson $R_\mathrm{AA}$ between the two different charm quark production channels. If we increase the gluon lifetime by hand, the final $D$ mesons from the gluon splitting channel become more suppressed at high $p_\mathrm{T}$ due to the larger energy loss experienced by hard gluons before splitting into heavy quark-antiquark pairs. We also note that although a single gluon does lose much more energy than a single $c$-quark within a given time, the gluon splits into a pair of $c\bar{c}$ in the end. As a result, the effective energy loss per heavy quark from the gluon-splitting process is not very different from that of heavy quarks produced by the initial hard scattering. This can be seen from the plot as well, i.e., a moderate variation of gluon lifetime does not significantly affect the final $D$ meson $R_\mathrm{AA}$. In Fig.~\ref{fig:plot-RAA_DD5-LHC-0-7d5-2}, we present the nuclear modification factor of $D$ mesons with the inclusion of both the direct and the gluon splitting processes, compared with the previous results without the contribution from gluon splitting. The error band in the figure corresponds to the variation of the gluon splitting time from the realistic value $\tau_\mathrm{split}$ extracted from \textsc{Pythia} to infinite value (meaning the gluon splitting into a heavy quark-antiquark pair happens outside the QGP medium). We can see that the contribution from gluon splitting to the nuclear modification of single $D$ meson spectra is very modest, even in the infinite $\tau_\mathrm{split}$ limit. We have verified that the variation of the starting time $\tau_0$ of the hydrodynamical evolution has little effect on the relative contributions from these two channels to the final suppression of $D$ mesons. As a final check, we also calculate the contribution of the gluon splitting to the single $B$ meson production in heavy-ion collisions and find that the influence of the gluon splitting channel on the $B$ meson $R_\mathrm{AA}$ is even smaller due to the shorter splitting time of the $g\rightarrow b\bar{b}$ process.


{\em Summary --} 
We have investigated the effect of the gluon-splitting process on the single inclusive heavy meson production in relativistic heavy-ion collisions. The splitting probability of hard gluons into heavy quark-antiquark pairs, the distribution of gluon splitting time, and the gluon to heavy quark-antiquark pair splitting function are extracted from \textsc{Pythia} simulations, which are then embedded into our Langevin transport model that incorporates both collisional and radiative energy loss for the hard partons propagating through a QGP matter. While the gluon splitting channel contributes quite a sizable fraction of the final $D$ meson production cross section, the influence on the final nuclear modification of the single heavy meson production is quite modest, due to the limited time for the hard gluons to interact with the dense medium before splitting into heavy quark-antiquark pairs.


{\em Acknowledgments --} 
We are grateful to discussions with R. Vogt and A. Dainese, and thank the computational resources provided by the Open Science Grid (OSG). S. Cao would like to thank S. Bass for valuable advice on establishing the heavy quark transport model at Duke University. This work is funded by the Director, Office of Energy Research, Office of High Energy and Nuclear Physics, Division of Nuclear Physics, of the U.S. Department of Energy under Contract No. DE-AC02-05CH11231, and within the framework of the JET Collaboration, by the Natural Science Foundation of China (NSFC) under Grant No. 11221504 and No. 11375072, by the Chinese Ministry of Science and Technology under Grant No. 2014DFG02050 and by the Major State Basic Research Development Program in China (No. 2014CB845404).

\bibliographystyle{h-physrev5}
\bibliography{SCrefs}

\end{document}